# Optimal double resonant condition in metallic core-shell nanocavity for third harmonic generation

Wenbo Zang,[1] Lingling Fan,[1] Xin Yang,[1] Mingyu Ma,[1] Peng Zhan,[1,2,3] Zhuo Chen,[1,2,*]
and Zhenlin Wang[1,2,4]

[1] *School of Physics and National Laboratory of Solid State Microstructures, Nanjing University, Nanjing 210093 China.*

[2] *Collaborative Innovation Center of Advanced Microstructures, Nanjing University, Nanjing 210093, China.*

[3]*zhanpeng@nju.edu.cn*

[4]*zlwang@nju.edu.cn*

**zchen@nju.edu.cn*

As the rapid development of nonlinear optics, the enhancement of optical third harmonic generation becomes a pop research realm in the fields of physics, chemistry, biology, materials science, information science and other fields. In this letter, theoretical analysis on double resonance situation in metallic core-shell nanostructure is performed in order to optimize the efficiency of third harmonic generation. As plenty of cavity modes ranging from visible area to near infrared can be excited efficiently in core-shell nanospheres, kinds of double resonant conditions can be formed by matching two different multipolar cavity modes. Numerical simulations show that the third harmonic generation (THG) intensity in the far field can be enhanced remarkably when the THG signal couples well with the high order cavity mode. More importantly, THG efficiency is optimum from the cavity on double resonant conditions coupling two modes with the same order. The THG intensity differs up to 3 magnitudes on this condition. Subsequent theoretical analysis indicates that changing third-order nonlinear susceptibility of the metal shell while keeping that of the core fixed has almost no effect on THG efficiency. This finding about optimal double resonance has an appreciable effect on optimizing THG in spherical cavity structures.

## I. INTRODUCTION

Nonlinear optical harmonic generation [1] is important in a broad range of technologies, such as photonics [2-4], material science [5] and biosensing [6,7]. Since the interactions between light waves in a nonlinear dielectric was researched initially in 1962 [8], researchers have made substantial and meaningful progress in nonlinear optics [9,10]. In particular, THG, a process that coherently triples the energy of the incident photon, can be induced regardless of symmetry of the media, in contrast to the case of second harmonic generation (SHG) [11]. In recent twenty years, the phenomenon of metal surface plasmon resonance (SPR) attracts many researchers [12]. It is acknowledged that when a local surface plasmon resonance (LSPR) of metal particles is excited, a strong enhancement of optical near-field can be observed [13]. Since the generation efficiency of nonlinear signals is positively related with the high order of electric field intensity excited by pump light, orders of magnitude enhancement of nonlinear effects can be achieved [14,15]. Though there are many works devoted to the enhancement of nonlinear harmonic generation in recent years, most of them concentrated on the optical response of nanostructures on single resonant situation [11,12,16-18], in which a cavity mode of the nanostructure is excited by the pump light. In these works, the purpose of adjusting nanostructures' geometric parameters is to optimize the resonance effects at pump wavelength. In order to achieve stronger nonlinear signals, a few people pay attention to double resonant conditions that the nanostructure resonates at the fundamental frequency and the triple (or double) frequency simultaneously [19-24]. Almost all of these works just simply illustrate that double resonance can increase or reduce [25] nonlinear signals. Rarely do works explore mechanism of double resonance's influence on nonlinear harmonic generation and how to achieve optimal results thoroughly. There are many cavity resonances which can be excited inside in spherical dielectric-metal core-shell nanocavities because of phase delay [26,27]. Meanwhile frequencies of different cavity modes can be changed easily by adjusting its geometric parameters [28] so we choose this nanostructure to investigate the influence of double resonance on THG.

In the process of SHG or THG in macroscopic nonlinear crystals, the pump light and corresponding harmonic light should fulfil the wave vector condition in order to achieve a greatest generation efficiency. The micro-mechanism of perfect wave vector condition is the nonlinear polarization radiation of every molecule's coherent superposition along harmonic wave vector. In the nonlinear process of single particle dominated by LSP, the radiation of nonlinear signal source still achieves coherent superposition in the far field, while the pump and harmonic lights have no traditional wave vector. So the final far-field harmonic signal can be optimal if appropriate LSP modes couple with the pump and harmonic light respectively.

In this letter, we simulated THG intensity on different double resonance situations in a metallic core-shell nanostructure whose core material has a high third order nonlinear susceptibility. Overly thick metal shell increases the loss of pump light illuminating in and THG signal emitted out, which all weaken the THG signal in far field. But if the metal shell is not thick enough, the plasmonic cavity modes have poor localization and large line width, which greatly reduces coupling between two resonant cavity modes excited by pump light and THG signal respectively. So we choose a moderate thickness, 50nm, of metal shell. Beyond that, we appropriately enlarge the radius of the nanocavity to 500nm for more cavity modes. The calculated results by COMSOL Multiphysics indicates that lots of cavity modes ranging from visible area to near infrared can be excited efficiently in core-shell nanospheres. The corresponding resonant wavelength can be tuned easily by various methods. So it is convenient to acquire a double resonant situation which means the pump light and the THG signal both can be coupled to some cavity modes. Numerical simulations show that the THG intensity in the far field can be enhanced remarkably when the THG signal couples well with the high order cavity mode. Also, the coupling efficiency depends on the multipolar moment of the THG signal which is governed by the resonant symmetry at the pump wavelength. Thus the symmetry matching of the involved cavity modes may heavily affect the THG efficiency. It is calculated that when double resonant situation occurs the THG intensity is greatly enhanced, compared to single resonant situation. Further, when the two cavity modes involved match well with each other, namely with the same multipolar moment, the final THG efficiency will display a dramatic enhancement up to 3 magnitudes. Subsequent theoretical analysis indicates that changing third-order nonlinear susceptibility of the metal shell while keeping that of the core constant has almost no effect on THG efficiency.

## II. CALCULATION DETAILS AND RESULTS

As a result of phase delay, the nanospheres have lots of plasmonic cavity modes ranging from visible area to near infrared in dielectric-metal core-shell nanocavity while the cavity modes all have narrow line width and different resonant frequencies, which can be adjusted easily by changing nanostructures' geometric parameters. So this nanostructure is very appropriate to explore what the effect of double resonant conditions is on THG. In this letter, our research object is a single dielectric-metal core-shell nanosphere in air. The dielectric core's refractive index is 1.59 with a radius of 500nm and the metal shell is silver with a thickness of 50nm. In order to facilitate the analysis, we suppose that the core has a high isotropic third order nonlinear susceptibility which contains only off-diagonal elements and ignore the shell's third nonlinear susceptibility.

Using the tabulated optical parameters of silver [29], we analyzed these plasmonic cavity modes of this Ag sphere cavity sample by simulating its linear optical response excited by linear polarization of the plane wave. Fig. 1(a) is total linear absorption spectrum numerically simulated by COMSOL Multiphysics. There are lots of plasmonic cavity modes ranging from visible area to near infrared excited efficiently in the cavity. We choose 1390nm, the resonant wavelength of TM21 mode in near infared, as pump wavelength in the next nonlinear simulation. And in the corresponding triple frequency range, five high order cavity modes, TM52 mode(472nm), TE13 mode(478nm), TM23 mode(487nm), TE61 mode(496nm) and TE32 mode(500nm), are used to achieve double resonant conditions later.

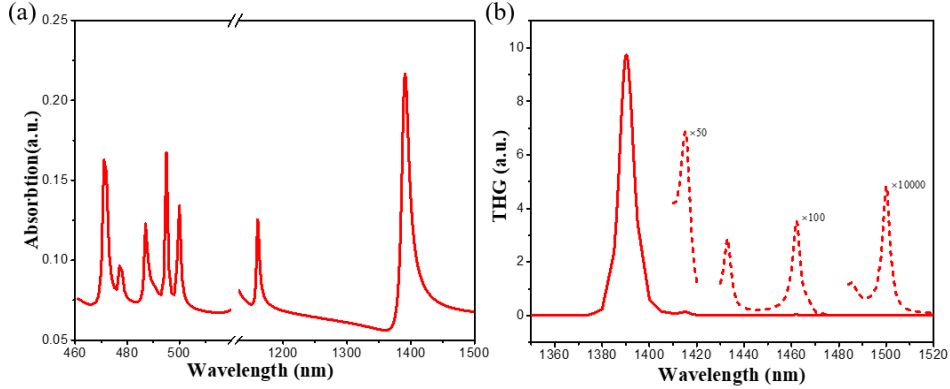

FIG. 1. Linear and nonlinear response of the plasmonic dielectric-metal core-shell nanocavity excited by plane wave. (a) Linear absorption spectrum numerically simulated by COMSOL Multiphysics. (b) THG intensity when the pump wavelength is varied from 1350nm to 1520nm in the near infrared. The pump light is coupled to TM21 mode at 1390nm.

For nonlinear simulations, third-harmonic fields are calculated by assuming that the fundamental fields are negligibly affected, undepleted pump approximation [30]. In other words, the fields at the fundamental wavelength are used in the source term for computing the response at the third-harmonic wavelength. The isotropic nonlinear material's third-order nonlinear is given by: $P = \varepsilon_0 \mathcal{X}^{(3)}(E \cdot E)E$ [31]. Here we take the high core's and ignored shell's third order nonlinear susceptibility $\mathcal{X}^{(3)}_{core} = 1 \times 10^{10}(m/V)^2$ and $\mathcal{X}^{(3)}_{shell} = 0(m/V)^2$ respectively. To explore the influence of double resonance on THG, we simulated the THG intensity from this nanosphere cavity excited by pump wavelength changed from 1350nm to 1520nm in near infrared, shown as Figure 1b. There are 6 distinguishing peaks in the THG efficiency spectroscopy. The one of them, located at the shortest wavelength and with a highest peak value, has a wavelength of 1390nm, which is precisely the resonant wavelength of TM2 cavity mode. Apparently the reason why the THG peak appears is that TM2 cavity resonance excited by pump light enhances electric field in the sphere cavity then the THG intensity. Beyond that, there are other 5 weaker peaks on the right of the spectroscopy. Compared to the linear spectroscopy in Fig. 1(a), we find that these 5 peaks' positions and the triple wavelength of the 5 high order cavity modes have one-to-one correspondence. In another word, THG signal can excite high order cavity modes in the cavity when pump wavelength is just the triple wavelength of them. We can conclude that both plasmonic cavity resonance excited by pump light and THG signal can enhance the THG intensity in far field.

To achieve double resonant conditions, we tune the refractive index of dielectric in visible range to an appropriate value which makes the resonant wavelength of a high order cavity be a third of pump wavelength. Taking the coupling of TM21 mode and TM52 mode as an example, we suppose that the refractive index of dielectric in near infrared is still 1.59 while that in visible is 1.56 because of dispersion. Then pump light of 1390nm wavelength excites TM21 mode and at the same time the THG signal excites TM52 mode. The same method of achieving double resonant conditions are used to couple TM21 with the other high order modes and the THG intensity on different conditions are shown in Fig. 2(a).

Comparing Fig. 1(b) and Fig. 2(a), the THG intensity emitted from the nanosphere on double resonant conditions is more significantly improved than that on not only single pump resonant situation but also single THG resonant situation. What's more, the THG intensity on the double resonant condition that TM21 mode is coupled well with TM23 displays a dramatic magnification up to 3 magnitudes, while the magnification of the other 4 double resonant conditions is about 1 magnitude. It is obvious that the double resonant condition achieved by the coupling of two cavity modes in the same order can lead to the most efficient enhancement of THG. In order to verify the above-mentioned conclusion, we repeat the simulation of THG efficiency on double resonant conditions, replacing TM21 mode with TE11 mode with 1166nm resonant wavelength. The result is shown as Fig. 2(b) and as expected we arrive the same conclusion that THG efficiency is optimum from the cavity on double resonant conditions coupling two modes with the same order.

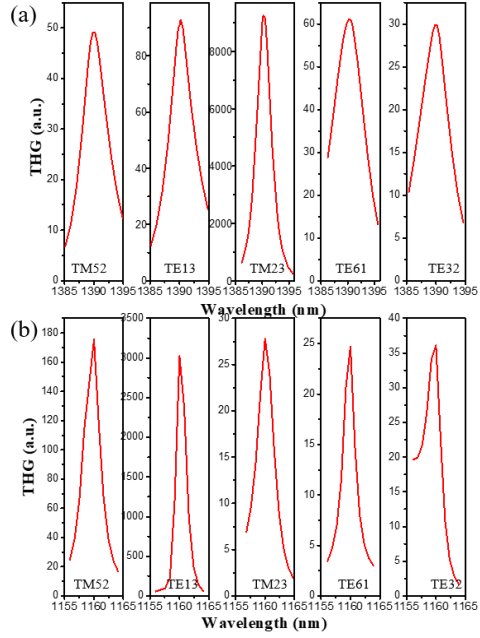

FIG. 2. THG intensity (a) at the pump wavelength of 1390nm(TM21) in different double resonant situations by tuning the refractive index of dielectric in visible area to make the THG signal coupled to the five high order cavity modes respectively when the pump light is coupled to TM2 mode and (b) at the pump wavelength of 1161nm(TE1) in different double resonant situations by tuning the refractive index of dielectric in visible area to make the THG signal coupled to the 5 high order cavity modes respectively when the pump light is coupled to TE1 mode.

### III. EXPLORATION AND DISCUSSION

For further exploration, we calculate the electric field distributions of the seven cavity modes we used above and THG's electric field distributions and three-dimensional far-field radiation patterns, shown as Fig. 3. The pump light propagates along the z axis and its electric field polarizes along the x axis. The first column (a) is the five high order modes' electric field distributions in k-E plane and the right side is that of the two fundamental modes, TM21 mode and TE11 mode in Fig. 3. Apparently the two pairs of modes with the same angular quantum number, TM21&TM23 and TE11&TE13, have extremely similar rotation symmetry respectively. And due to their same rotation symmetry, the two pairs of modes have a good overlap electric field distribution in space. From third-order nonlinear polarization formula, we infer that the good overlap of electric field distribution between same-order pump and THG modes results a stronger nonlinear source and the THG is enhanced in generation process.

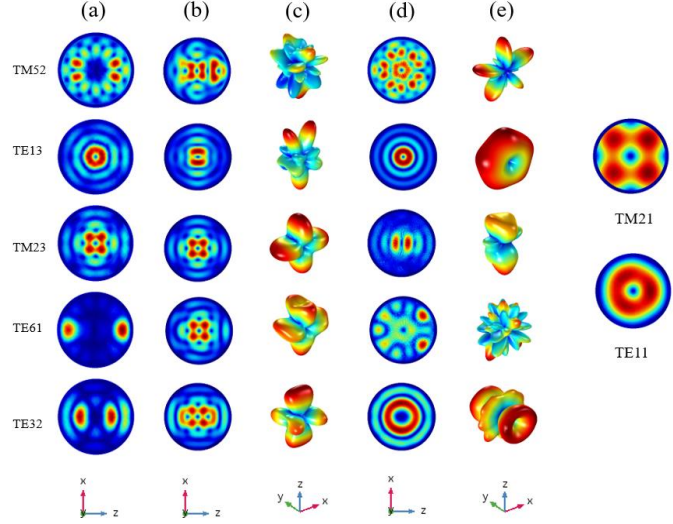

FIG. 3 (a) The electric field distributions in k-E plane of TM52 mode, TE13 mode, TM23 mode, TE61 mode and TE32 mode. (b, d) The electric field distributions in k-E plane of THG when the five high order cavity modes are coupled with TM21 mode(b) and TE11 mode(d) respectively. (c, e) 3D patterns of the far-field TH radiation when the five high order cavity modes are coupled with TM21 mode(c) and TE11 mode(e) respectively.

Fig. 3 (b, d) are the electric field distributions in k-E plane of THG when the five high order cavity modes are coupled with TM21 mode (b) and TE11 mode (d) respectively and their corresponding 3D patterns of the far-field TH radiation are shown in Fig. 3 (c, e). It is obvious that THG on the same-order double condition has an almost same electric field distribution with the cavity mode at TH wavelength. As we know, the cavity mode's symmetry of current wavelength determines THG's far-field radiation in space. When there are a number of coherent point sources in the nanosphere, the point sources' intensity and orientation have strong influence on the coupling efficiency between the source and the radiant cavity. The optimal radiation efficiency can be achieved if total electromagnetic moment of the coherent point sources coincides with that of the resonant cavity mode. In the same way, the highest coincidence of symmetry between the two modes with the same angular momentum lead to the strongest THG emitted from the sphere cavity.

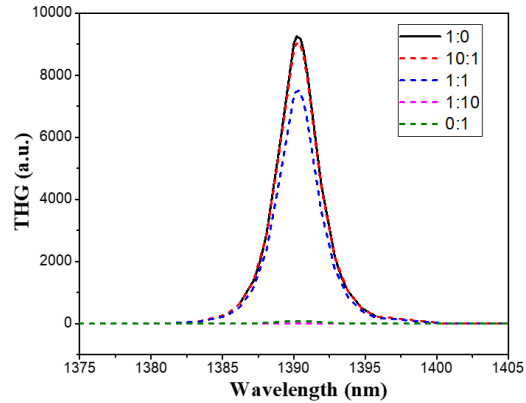

FIG. 4 The THG intensity from the dielectric-metal core-shell nanocavities with different third-order nonlinear susceptibility ratio between the core's and the shell's.

In our simulation by COMSOL Multiphysics, we suppose that the core has a high third order nonlinear susceptibility and ignore the shell's third nonlinear susceptibility. But in fact the metal commonly has an isotropic high nonlinear susceptibility, and thus it's challenging for researchers to find a kind of dielectric with higher third order nonlinear susceptibility for experimental verification. Changing the third-order nonlinear susceptibility of dielectric core and that of metal shell respectively, we calculate the THG intensity on TM21-TM23 double resonant conditions with different ratio between them shown as Fig. 4 for feasibility experiment in the next. Calculation results indicates that third-order nonlinear susceptibility ratio between the core's and the shell's has little effect on THG in far field. We guess that the slightly reduced THG intensity (blue short dash in Fig. 4), when the shell has the same high third nonlinear susceptibility with the core, is due to destructive interference between the two parts THG form the core and shell. In other words, a higher third order nonlinear susceptibility of the core is not strictly required in real experiments and we can use any dielectric as the core's material without considering the metal's nonlinear susceptibility, which makes the experimental verification much easier.

## V. CONCLUSION

In summary, we simulated THG intensity on different double resonance situations in a metallic core-shell nanostructure whose core material has a high third order nonlinear susceptibility. The corresponding resonant wavelength can be tuned easily by various methods, for example, tuning the refractive index of dielectric or its geometric parameters. So it is convenient to acquire any double resonant situation we want. Numerical simulations show that the THG intensity in the far field can be enhanced remarkably while the THG signal couples well with the high order cavity mode. And the coupling efficiency depends on the multipolar moment of the THG signal which is governed by the resonant symmetry at the pump wavelength. Thus the symmetry matching of the involved cavity modes may greatly affect the THG efficiency. It is calculated that when double resonant situation occurs, the THG intensity is enhanced more than single resonant situation. And when the two cavity modes involved match well with each other, the final THG intensity will engage a dramatic enhancement up to 3 magnitudes. Subsequent theoretical analysis indicates that varying third-order nonlinear susceptibility of the metal shell and fixing that of the core have almost no effect on THG.